\documentclass[conference]{IEEEtran}
\IEEEoverridecommandlockouts
\usepackage{cite}
\usepackage{amsmath,amssymb,amsfonts}
\usepackage{algorithmic}
\usepackage{graphicx}
\usepackage{textcomp}
\usepackage{xcolor}
\usepackage{subcaption}
\usepackage{hyperref}

\usepackage{booktabs}
\usepackage{makecell}
\usepackage{fontawesome5}

\usepackage{listings}
\usepackage{xcolor}
\def\BibTeX{{\rm B\kern-.05em{\sc i\kern-.025em b}\kern-.08em
    T\kern-.1667em\lower.7ex\hbox{E}\kern-.125emX}}

\usepackage{booktabs}
\usepackage{caption}
\usepackage{subcaption}
\usepackage{array}
\newcolumntype{P}[1]{>{\centering\arraybackslash}p{#1}}

\begin{document}

\title{Making Sense of Scams: Understanding Scam Conversations Through Multi-Level Alignment}

\author{Zhenyu Mao$^{1}$, Jacky Keung$^{1}$, Xiangyu Li$^{2}$, Yicheng Sun$^{1}$, Kehui Chen$^{1}$, Jingyu Zhang$^{3,*}$, and Jialong Li$^{4}$\\
	\normalsize $^{1}$City University of Hong Kong, Hong Kong, China $^{2}$SeysoAI, Suzhou, China\\
    \normalsize $^{3}$Hong Kong Metropolitan University, Hong Kong, China $^{4}$Waseda University, Tokyo, Japan\\
	\normalsize zhenyumao2-c@my.cityu.edu.hk, Jacky.Keung@cityu.edu.hk, xiangyu.li@seysoai.com \\
    \normalsize yicsun2-c@my.cityu.edu.hk, kehuichen2-c@my.cityu.edu.hk, lijialong@fuji.waseda.jp\\
	\normalsize *corresponding author: fzhang@hkmu.edu.hk 
}

\maketitle

\begin{abstract}

Online scams often unfold gradually through interaction, yet existing detection systems predominantly rely on snapshot-based signals and interruptive warnings, revealing two research gaps in the lack of signals that represent scam risk within conversational dynamics and the underexplored design of non-interruptive interaction.
To address these gaps, we introduce multi-level alignment-based hints, informed by the Interactive Alignment Model, as a new detection signal for supporting sensemaking in scam-related conversations.
These hints operationalize low-level lexical and syntactic alignments and high-level semantic and situation-model alignments between conversational participants, making conversational dynamics visible to users.
We first conduct a preliminary evaluation on real-life scam dialogues, showing that as conversations approach scam attempts, low-level alignment scores remain stable while high-level alignment scores systematically decline, revealing a consistent cross-level pattern indicative of scam progression.
Building on this insight, we conduct a user study with thirty participants, indicating that relative to the no-hint baseline, multi-level alignment-based hints increase precision by 0.25, recall by 0.16, and F1 score by 0.21, yielding substantially larger gains than the marginal improvements achieved by keyword-triggered alerts.
Statistical analyses reveal that the proposed hints support earlier and more stable confidence formation over time, with ablation results further highlighting the effectiveness of combining alignment hints across levels in achieving these advantages.

\end{abstract}


\section{Introduction}
\label{sec:introduction}

Online scams are a growing source of financial and psychological harm, with recent academic research documenting their impact on victims' well-being, trust, and social relationships across online platforms \cite{anderson2013measuring}.
Prior studies show that many contemporary scams, such as romance or investment scams, no longer rely on single deceptive messages, but instead unfold through prolonged interpersonal interaction in which deceptive intent emerges gradually \cite{oak2025victims}.
A dominant paradigm in Human–Computer Interaction (HCI) research on scams treats scam detection as a primarily technical problem, with user interaction considered only after a detection has been done.
In this paradigm, detection systems identify snapshot-based signals of malicious activity, while HCI research examines how users perceive and respond to the resulting warnings \cite{sunshine2009crying, akhawe2013alice}.
For example, a systematization of phishing user studies characterizes a common approach in which systems detect risk in a snapshot-based manner and communicate it to users through interruptive warnings \cite{das2019sok}.

Key limitations of this paradigm are that snapshot-based detection systems struggle to capture how risk emerges through conversational dynamics (i.e., how linguistic and conceptual coordination between participants evolves over the course of interaction \cite{pickering2006alignment}), and that interruptive warnings offer users limited support for understanding or reasoning about risk.
This reveals two research gaps: first, the lack of detection signals that represent scam risk within conversational dynamics as an evolving process rather than isolated events, and second, the limited support for user sensemaking within existing interaction designs, which often substitute automated judgment for user interpretation rather than scaffolding it.
Addressing these gaps requires rethinking both the detection signals and their integration: developing process-oriented representations of conversational risk while embedding them in ways that scaffold user sensemaking rather than interruption \cite{akhawe2013alice}.

To bridge these gaps, we draw on the Interactive Alignment Model, which explains how dialogue participants build shared understanding through alignment across multiple representational levels \cite{pickering2004toward, pickering2006alignment}.
Because alignment provides a theoretically grounded lens on how conversational coordination develops and deviates over time, we operationalize multi-level alignment-based hints that surface coordination between participants at both low levels (lexical and syntactic alignments) and high levels (semantic and situation-model alignments).
The proposed hints combine a process-oriented perspective that models scam risk as an evolving interactional phenomenon, and a sensemaking-oriented interaction design that integrates risk cues without disrupting the communication.

This work makes three primary contributions.
First, we operationalize multi-level alignment to support human sensemaking of conversational dynamics, and validate the effectiveness of alignment-based hints for scam detection. 
Second, we conduct a preliminary evaluation on real-life scam dialogues, revealing a consistent cross-level pattern in which high-level alignment declines while low-level alignment remains comparatively stable as scam attempts approach. 
Third, we invited thirty participants for a user study, and both objective and subjective data demonstrate that multi-level alignment-based hints improve scam detection accuracy and support earlier and more stable confidence formation compared with keyword-triggered alerts, with ablation analysis highlighting the effectiveness of combining alignment hints across levels.

\begin{figure*}[!t]
    \centering
    \includegraphics[width=140mm]{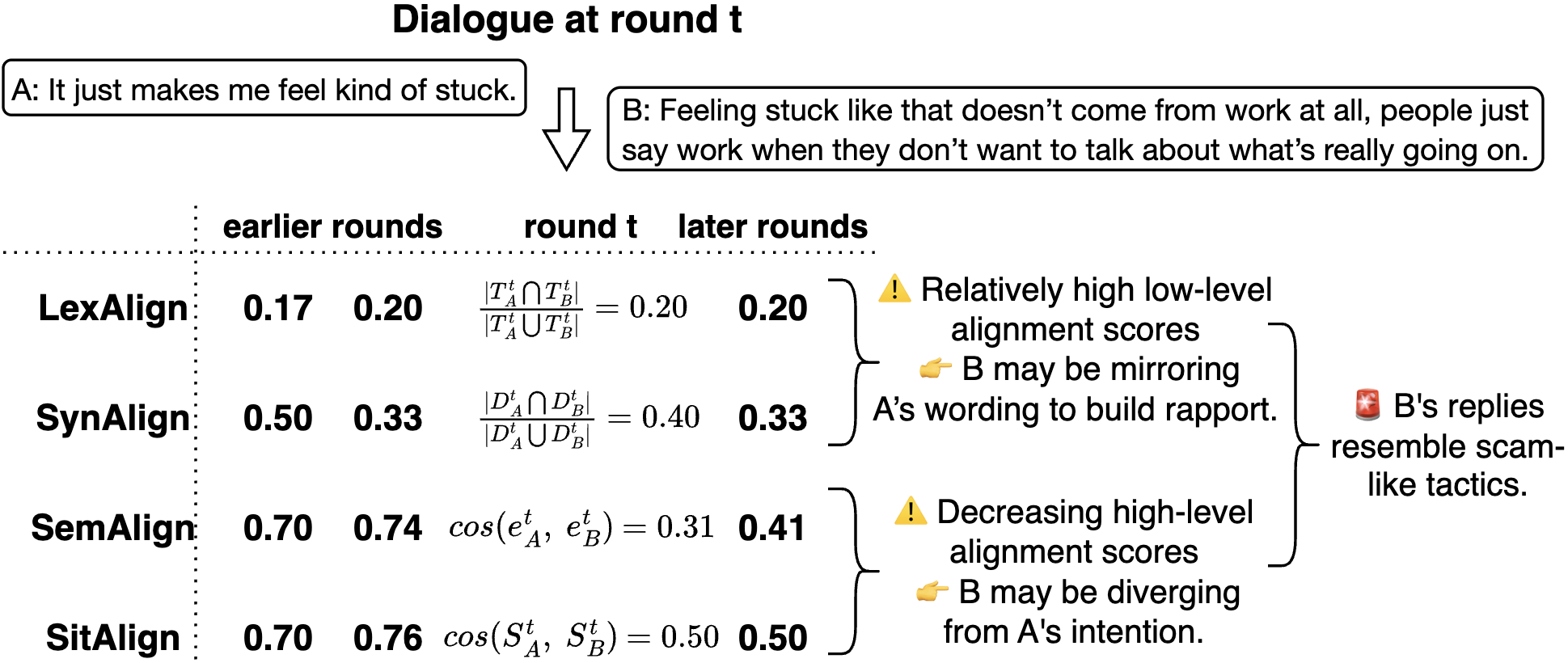}
    \caption{Alignment analysis on example dialogue round}
    \label{fig:example}
\end{figure*}
\section{Background and Related Work}
\label{sec:background}

\subsection{HCI in Scam-Related Scenarios}

Recent HCI work on mitigating scams largely continues within the established interaction paradigm.
\cite{tamal2024unveiling} proposed an enhanced feature vectorization and supervised learning approach for phishing detection, where static linguistic features are aggregated into a classifier output.
\cite{buono2023let} proposed novel warning dialogs for email clients that identify phishing emails based on automated detection at the moment a user attempts to open a suspicious email.
\cite{desolda2023explanations} discussed phishing detection systems that trigger a warning dialog when an automated backend flags a visited website or email as a potential phishing threat, based on static classification and feature detection.
Across these systems, detection relies on snapshot-based signals computed at isolated points in time, and detected risk is communicated through interruptive warnings.
While effective for blocking known threats, this paradigm offers limited support for users’ sensemaking of how risk unfolds and evolves within conversations.

\subsection{Interactive Alignment Model}

The Interactive Alignment Model provides a psycholinguistic view of how conversational participants achieve mutual understanding through coordination across multiple levels of language and meaning, including lexical, syntactic, semantic, and situation-level representations \cite{pickering2004toward, pickering2006alignment}.
Rather than relying on explicit reasoning about a partner's mental state, the model suggests that successful conversation often emerges through automatic alignment processes, whereby one speaker's linguistic choices make similar choices more likely for the other speaker.
Prior empirical work has examined alignment at individual levels of communication, such as lexical \cite{schober1989understanding}, syntactic \cite{branigan2000syntactic}, and shared situation models \cite{garrod2009joint}.
These studies show that alignment at individual levels supports efficient interaction, while how alignment across levels (e.g., syntactic and semantic alignments) jointly shapes users' interpretation, reasoning, or trust during interaction remains unclear.
\section{Multi-Level Alignment-Based Hints}
\label{sec:alignment}

We operationalize multi-level alignment-based hints using four alignment signals, including lexical, syntactic, semantic, and situation-model alignment, computed at the level of dialogue rounds and grouped into low-level and high-level categories \cite{pickering2004toward}.
Lexical and syntactic alignment are treated as low-level because they reflect local, often automatic coordination in wording and structure, whereas semantic and situation-model alignment are treated as high-level because they require integration of meaning and discourse context across turns, reflecting deeper convergence in interpretation and intent \cite{pickering2004toward}.

\paragraph{Dialogue Rounds}
We model each dialogue as an ordered sequence of messages following an $A \rightarrow B \rightarrow A \rightarrow B \rightarrow \dots$ structure.
A dialogue round $t$ is defined as an ordered adjacent pair $(u_t^A, u_t^B)$, where speaker $A$ produces an initiating message $u_t^A$ and speaker $B$ produces a response $u_t^B$.
Alignment is computed once per round on this ordered $A$--$B$ pair, reflecting the asymmetric roles in scam conversations and the theoretical assumption that scammers strategically adapt their language to build rapport and trust with victims.

\paragraph{Lexical Alignment}
Lexical alignment captures overlap in content-word usage between the two turns in a round. Let $T_A^t$ and $T_B^t$ denote the sets of unique content words (after lowercasing and stopword removal) in $u_t^A$ and $u_t^B$, respectively. Lexical alignment at round $t$ is computed using Jaccard similarity as $\text{LexAlign}_t = |T_A^t \cap T_B^t| / |T_A^t \cup T_B^t|$.

\paragraph{Syntactic Alignment}
Syntactic alignment captures similarity in grammatical structure. Let $D_A^t$ and $D_B^t$ denote the sets of dependency relation labels extracted from $u_t^A$ and $u_t^B$, respectively. Syntactic alignment at round $t$ is computed as the Jaccard similarity $\text{SynAlign}_t = |D_A^t \cap D_B^t| / |D_A^t \cup D_B^t|$.

\paragraph{Semantic Alignment}
Semantic alignment captures similarity in utterance-level meaning. Let $e_A^t$ and $e_B^t$ denote the sentence-level embeddings of $u_t^A$ and $u_t^B$, respectively, obtained from a pretrained sentence encoder and $\ell_2$-normalized. Semantic alignment at round $t$ is computed as the cosine similarity $\text{SemAlign}_t = \cos(e_A^t, e_B^t)$.

\paragraph{Situation-Model Alignment}
Situation-model alignment captures similarity in interlocutors' evolving discourse states across rounds. For each speaker $p \in \{A,B\}$, we maintain an exponentially smoothed discourse state $S_t^p$, initialized as $S_0^p = \mathbf{0}$ and updated at each round as $S_t^p = \alpha S_{t-1}^p + (1-\alpha)e_p^t$, where $e_p^t$ is the embedding of speaker $p$'s utterance at round $t$ and $\alpha \in [0,1]$ controls the influence of prior context. Situation-model alignment at round $t$ is then computed as $\text{SitAlign}_t = \cos(S_t^A, S_t^B)$.

According to the Interactive Alignment Model \cite{pickering2004toward}, in genuine interactions, these levels tend to co-evolve toward mutual understanding.
In contrast, scam conversations often exhibit asymmetric or strategically managed alignment dynamics, e.g., stable or increasing low-level alignment scores combined with decreasing high-level alignment scores suggest surface-level linguistic accommodation without corresponding convergence in meaning or discourse state.
We operationalize both the alignment scores and patterns in their dynamics as non-interruptive sensemaking hints, enabling users themselves to assess whether observed coordination reflects genuine shared understanding or merely surface-level adaptation consistent with manipulative rapport-building, as illustrated in Figure~\ref{fig:example}.
\section{Evaluation}
\label{sec:evaluation}

Research questions (RQs) are set as follows.

\begin{itemize}
    \item \textbf{RQ1 (Pattern identification).}
    Do real-world scam dialogues exhibit a pattern of decreasing high-level alignment scores alongside stable low-level alignment scores, as hypothesized in Section~\ref{sec:alignment}?
    \item \textbf{RQ2.1 (Objective effectiveness).}
    To what extent do multi-level alignment-based hints improve users’ scam detection performance, as measured by precision, recall, and F1 score?
    \item \textbf{RQ2.2 (Subjective effectiveness).}
    How do multi-level alignment-based hints influence participants’ subjective confidence formation over time during scam detection?
\end{itemize}

\begin{figure}[hbtp]
    \centering
    \includegraphics[width=0.8\linewidth]{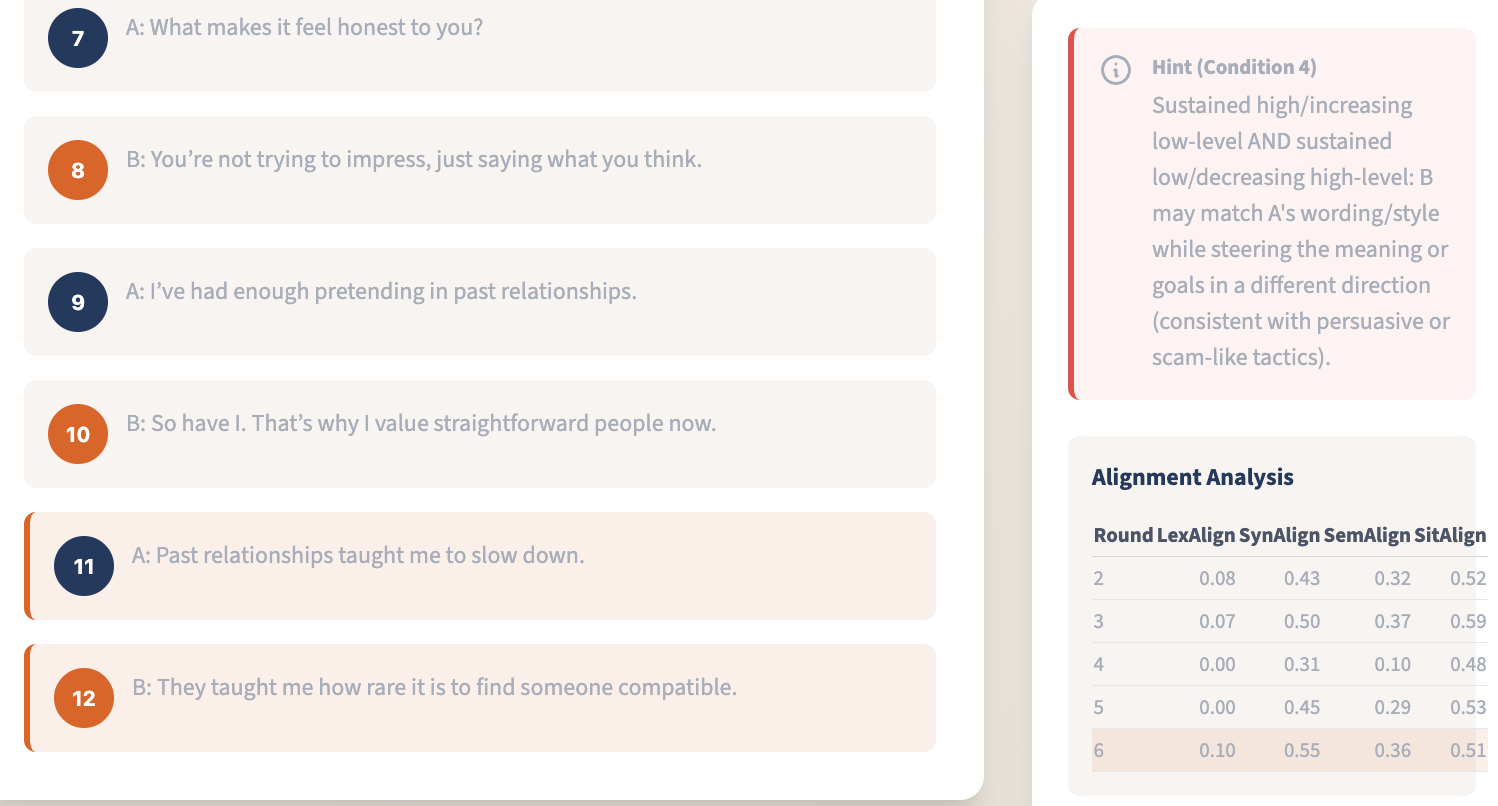}
    \caption{User study UI. Left: ongoing dialogue. Upper right: pattern hint. Lower right: alignment scores across rounds.}
    \label{fig:ui}
\end{figure}

RQ1 is evaluated through a preliminary analysis of alignment score dynamics in real-world scam dialogues from \cite{faber2024lovefraud02} dataset (explained in Section \ref{sec:dataset}).
RQ2.1 and RQ2.2 are investigated via a user study in which participants attempt to detect scams in the same dataset (Figure \ref{fig:ui}).

\subsection{Experiment 1 (Exp1): Preliminary analysis}
\label{sec:exp1}
\subsubsection{Exp1-Settings}

\paragraph{Dataset selection}
\label{sec:dataset}

The LoveFraud02 dataset \cite{faber2024lovefraud02} is a collection of real-world online scam conversations, providing a representative view of how conversational scams unfold from rapport building to scam attempts.
First, it reflects diverse trust-building personas and strategies, including romance pursuers, self-identified military officers, overseas professionals, and other fabricated identities to establish rapport over interaction.
Second, it documents a range of scam execution strategies, in which instrumental requests are gradually introduced, such as asking victims to purchase gift cards, temporarily hold or transfer money on the scammer’s behalf, or pay fees and margins associated with purported overseas employment.

\paragraph{Preprocessing}

Preprocessing was performed to ensure normalized and comparable alignment score calculations across conversations.
This step is necessary for two reasons: 1) conversations in the original dataset vary substantially in length, making direct comparison of alignment score trajectories difficult, and 2) the rapport-building phase in scam conversations can be repetitive, with similar interactional features recurring.
To enable meaningful comparison while preserving the core dynamics of rapport construction, we therefore focus on the dialogue rounds immediately before the point at which an explicit scam occurs, defined as the first message where the scammer raises a request directly related to financial matters.

For each conversation, we first manually identified the message in which the scammer made a financially related request for the first time.
All dialogue occurring after this point was excluded from subsequent analysis.
Conversations that did not contain any financially related request, or that only involved non-financial behaviors (e.g., requests for photos or suggestions to switch chat platforms), were excluded.
We then extracted the final 40 dialogue rounds (i.e., 80 messages) before the identified request.
To ensure consistent round-based alignment score computations, consecutive messages sent by the same participant were merged, resulting in an alternating speaker structure.
Conversations that did not contain at least 40 dialogue rounds prior to the request were excluded from subsequent analysis.
The original dataset contains 83 conversations, with an average of 637.67 messages per conversation.
After preprocessing, the dataset used for analysis consists of 47 conversations, each comprising exactly 80 messages.

\subsubsection{Exp1-Results}

\begin{figure}[hbtp]
    \centering
    \includegraphics[width=0.8\linewidth]{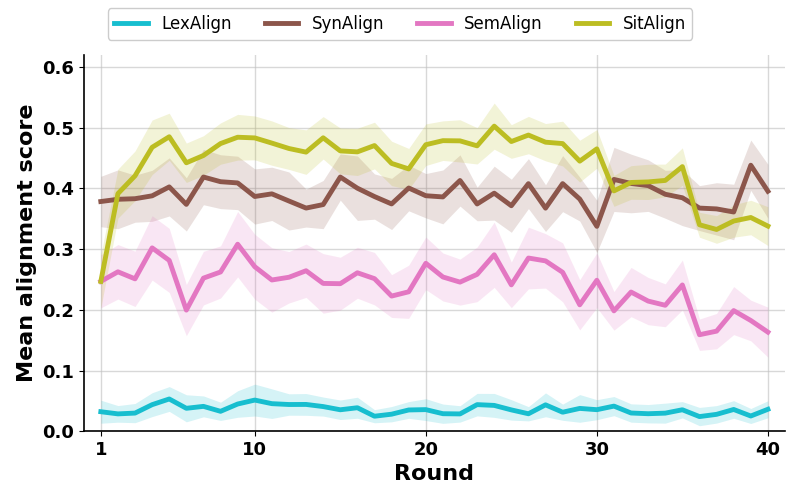}
    \caption{Mean Alignment Score Change per Dialogue Round with 95\% Confidence Interval}
    \label{fig:preliminary}
\end{figure}

As shown in Figure~\ref{fig:preliminary}, the trajectories of the four alignment scores from Round 1 to Round 40, where the scam attempt occurs, exhibit distinct patterns across representational levels.
During the earlier stage of interaction, alignment scores remain relatively stable across all levels, reflecting sustained conversational coordination during rapport building.
Lexical alignment score consistently ranges between 0.03 and 0.05, while syntactic alignment score remains around 0.38 throughout most of the dialogue.

In contrast, high-level alignment scores display a markedly different temporal pattern.
Semantic alignment and situation-model alignment scores remain relatively stable during the first 30 rounds, but both decline steadily over the final 10 rounds leading up to the scam attempt.
It is worth noting that situation-model alignment scores are lower in the earliest rounds by design, as this measure is computed using an accumulative formulation initialized from scratch (Section~\ref{sec:alignment}), however, the late-stage decline remains pronounced.

To formally assess temporal change, we computed per-conversation Spearman correlations between alignment scores and dialogue rounds and evaluated these using two-sided Wilcoxon signed-rank tests ($n = 47$).
The null hypothesis ($H_0$) assumes that the median per-conversation correlation equals zero, indicating no systematic monotonic temporal change.
Results indicate no significant monotonic change for lexical or syntactic alignment ($p > 0.05$), whereas both semantic and situation-model alignment exhibit significant negative monotonic trends ($p < 0.001$).

\begin{center}
\vspace{-4mm}
    \resizebox{\linewidth}{!}{
\begin{tabular}{l!{\vrule width 1pt}p{0.9\columnwidth}}
    \makecell{{\LARGE \faLightbulb}}  &\textbf{Finding 1.}
    During the earlier rapport-building stage, alignment scores are broadly stable across all levels. As the conversation approaches an explicit scam attempt, however, scammers tend to maintain low-level linguistic alignment scores to preserve surface rapport, while simultaneously shifting the interaction at deeper semantic and situational levels, as evidenced by the consistent decline in high-level alignment scores. This divergence between alignment scores emerges reliably across conversations and characterizes the transition from rapport maintenance to scam execution, and provides empirical support for the hypothesis in Section \ref{sec:alignment}.
\end{tabular}}
\end{center} 

\subsubsection{Exp1-Illustrative Case Study}

The following example presents five consecutive dialogue rounds around round 32, which marks a clear shift in alignment score dynamics, where semantic and situation-model alignment scores decrease noticeably while lexical and syntactic alignment scores remain stable.
Minor spelling and capitalization inconsistencies are preserved from the original data.

{\small
\noindent\textbf{Victim:} I am also exhausted. Today was very long with a lot of things to do. \\
\textbf{Scammer:} I'm so sorry about that, Hope everything is okay with you over there. \\
\textbf{Victim:} Everything is fine.... just some days are more work than others. \\
\textbf{Scammer:} Alright, Well I'm glad you are okay, I have good news for you and I can't wait to share it with you. \\
\textbf{Victim:} Tell me. \\
\textbf{Scammer:} Are you sure? \\
\textbf{Victim:} Why wouldn't I be? Good news is always nice to hear. \\
\textbf{Scammer:} Yeah of course. I wrote a request to the United Nations last week and I was informed that the release of my certificate of retirement will be a few weeks from now, Just yesterday Afternoon I was mailed by the Chairman chief of staff notifying me concerning my request that I'm approved of my arrival process and that I should go ahead to obtain a leave form that will replace and assume a new chain of command. The process is quick and I must obtain a Leave replacement form to warrant my change of command in terms of retirement. \\
\textbf{Victim:} Then everything sounds very positive. \\
\textbf{Scammer:} Yes my lovely Queen Pamela. This was the procedure regarding my Arrival and chain of command. firstly I requested a Letter of Approval which was Issued to me by the chief of staff. I'm so happy and I can't wait to be with you.
}

In this sequence, the scammer mirrors the victim’s concern and maintains similar conversational structure, resulting in sustained lexical and syntactic alignment.
However, a divergence emerges at higher representational levels when the scammer introduces a dense institutional and procedural narrative involving formal authorities, documentation, and administrative processes.
While the victim responds positively, their reply does not engage with the introduced details, indicating limited shared grounding of meaning.
This example illustrates how high-level alignment scores can decrease even when low-level linguistic alignment scores remain intact, particularly as the interaction approaches a scam attempt.

\subsection{Experiment 2 (Exp2): User Study}
\subsubsection{Exp2-Settings}

\paragraph{Dataset and Task}

The dataset consists of two balanced subsets.
The first subset includes 10 scam dialogues covering a range of common online conversational scams, drawn from \cite{faber2024lovefraud02}.
These scam dialogues were preprocessed to ensure consistency in length and style.
The second subset comprises 10 non-scam dialogues that closely mirror the scam dialogues in topic, length, style, and interaction format, while excluding deceptive or manipulative intent.
Each dialogue consists of 20 rounds, corresponding to 40 messages exchanged between two speakers.
Participants were asked to determine whether a given dialogue involved a scam or appeared to be evolving into a scam.
Dialogues were presented incrementally: participants could make a decision as soon as they felt confident based on the early rounds, or continue reviewing later rounds before giving a final judgment.

\paragraph{Hint Conditions}
We included five hint conditions that differed in the type and level of conversational support provided: the proposed multi-level alignment-based hint, two baseline conditions, and two ablation variants.
These hint conditions were counterbalanced across dialogues, and dialogues appeared in randomized order such that each participant experienced all five conditions an equal number of times.
In the alignment-based hint conditions, participants were shown the alignment scores for the five most recent rounds, together with a textual hint based on the pattern validated in Section \ref{sec:exp1}, as shown in Figure \ref{fig:ui}.
\begin{itemize}
    \item \textbf{Hint 0: No Hint}, showing no additional information.
    \item \textbf{Hint 1: Keyword-Triggered Alerts}, highlighting the presence of words commonly associated with scams.
    \item \textbf{Hint 2: Low-Level Alignment-Based}, only showing lexical and syntactic alignment scores.
    \item \textbf{Hint 3: High-Level Alignment-Based}, only showing semantic and situation-model alignment.
    \item \textbf{Hint 4: Multi-Level Alignment-Based}, showing all four alignment scores.
\end{itemize}

\paragraph{Metrics}
We collected both objective and subjective effectiveness measures throughout the task.
For the objective effectiveness, it is based on participants’ final judgments (\emph{scam} or \emph{non-scam}) for each dialogue, we computed precision, recall, and F1 score to evaluate performance under each hint condition.
For the subjective effectiveness, participants reported a confidence score on a 10-point scale at each round, where 1 indicated insufficient information to make a judgment and 10 indicated sufficient information.
As different hints provide supplementary cues that support participants in transitioning from insufficient to sufficient information for judgment, we analyzed confidence trajectories to assess how different hint conditions shaped confidence over rounds.

\paragraph{Participants}
Thirty native English speakers participated in the user study (19 male, 11 female; 7 aged 20–30, 20 aged 30–40, and 3 aged 40–50). The study collected no personally identifiable information, and was conducted in accordance with institutional guidelines for research involving human participants.

\subsubsection{Exp2-Results and Discussions}

\begin{figure}[hbtp]
\centering
\includegraphics[width=0.8\linewidth]{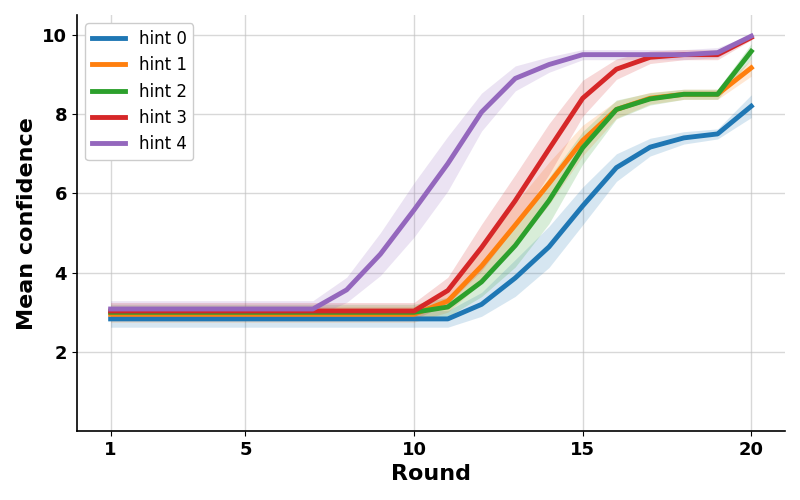}
\caption{Confidence Change per Round with 95\% Confidence Interval Reported by Participants}
\label{fig:data}
\end{figure}

Table~\ref{tab:data} summarizes participants' decision accuracy across hint conditions, while Figure~\ref{fig:data} compares changes in confidence trajectories over rounds across hint conditions.

\begin{table}[t]
  \centering
  \small
  \caption{Precision, recall, and F1 score across different hint conditions.}
  \label{tab:data}
  \begin{tabular}{c|ccc}
    \toprule
    \textbf{Hint Conditions} & \textbf{Precision} & \textbf{Recall} & \textbf{F1 Score} \\
    \midrule
    Hint 0 & 0.56 & 0.67 & 0.61 \\
    Hint 1 & 0.59 & 0.67 & 0.63 \\
    Hint 2 & 0.62 & 0.77 & 0.69 \\
    Hint 3 & 0.64 & 0.77 & 0.70 \\
    Hint 4 & \textbf{0.81} & \textbf{0.83} & \textbf{0.82} \\
    \bottomrule
  \end{tabular}
\end{table}

\paragraph{Results for RQ2.1}
\label{sec:exp1-results}

As shown in Table~\ref{tab:data}, keyword-triggered alerts yield only marginal improvements in objective detection performance relative to the no-hint baseline.
Precision increases slightly from 0.56 to 0.59 and F1 score from 0.61 to 0.63, while recall remains unchanged at 0.67.
In contrast, the proposed multi-level alignment-based hints produce substantial performance gains across all objective metrics.
Relative to the no-hint condition, precision increases by 25 percentage points (0.56 to 0.81), recall by 16 points (0.67 to 0.83), and F1 score by 21 points (0.61 to 0.82).
Ablation results further indicate that both low-level and high-level alignment-based hints improve detection accuracy relative to the no-hint baseline, but neither achieves the same level of performance as the multi-level condition.

To assess statistical reliability, we computed precision, recall, and F1 separately for each participant under each hint condition, and conducted Friedman tests across the five hint conditions, followed by post-hoc paired Wilcoxon signed-rank tests ($n = 30$).
The null hypothesis ($H_0$) assumes that detection performance does not differ across hint conditions.
No significant overall condition effects were observed for precision, recall, or F1 ($p > 0.05$), and pairwise comparisons likewise did not reach statistical significance.
This may reflect the limited number of trials per condition and the discrete, bounded nature of precision, recall, and F1 scores, which introduce tied observations and reduce test sensitivity.

\begin{center}
\vspace{-4mm}
    \resizebox{\linewidth}{!}{
\begin{tabular}{l!{\vrule width 1pt}p{0.9\columnwidth}}
    \makecell{{\LARGE \faLightbulb}}  &\textbf{Finding 2.}
    Relative to the no-hint condition, multi-level alignment-based hints yield higher aggregate detection performance than keyword-triggered alerts, yet further validation with larger sample sizes is needed.
\end{tabular}}
\end{center} 

\paragraph{Results for RQ2.2}

As shown in Figure~\ref{fig:data}, confidence trajectories differ markedly across hint conditions.
Compared with the no-hint baseline, keyword-triggered alerts exhibit only a minor effect on confidence formation.
Confidence begins to increase at a similar point in the dialogue (around Round 11) and follows a comparable growth pattern, reaching a plateau around Round 17.
The primary difference lies in a small increase in final confidence, rising from approximately 8 under no hint to approximately 9 under keyword-triggered alerts.
In contrast, participants exposed to multi-level alignment-based hints exhibit a different confidence trajectory. 
Confidence begins to rise several rounds earlier (around Round 8), increases more steeply once growth begins, and converges more quickly, reaching a stable plateau around Round 13.
Moreover, final confidence under the multi-level condition converges to a higher level, approaching the maximum of the confidence scale.
Ablation results indicate that single-level alignment-based hints primarily affect final confidence levels but do not produce the earlier onset or faster convergence observed under the multi-level condition.

To assess statistical reliability, we aggregated confidence separately for each participant under each hint condition and applied the same Friedman and Wilcoxon tests ($n = 30$) as in Section \ref{sec:exp1-results}.
Confidence differed significantly across conditions ($p < 0.001$), with the multi-level condition outperforming all baselines ($p < 0.001$).

\begin{center}
\vspace{-4mm}
    \resizebox{\linewidth}{!}{
\begin{tabular}{l!{\vrule width 1pt}p{0.9\columnwidth}}
    \makecell{{\LARGE \faLightbulb}}  &\textbf{Finding 3.}
    Multi-level alignment-based hints yield significantly higher overall confidence than all baseline conditions. While single-level alignment hints provide moderate improvements, combining hints across levels gives the strongest and most reliable confidence gains.
\end{tabular}}
\end{center} 

\paragraph{Discussion on Alignment-Based Conditions}

The moderate improvements observed under the single-level alignment conditions suggest that partial alignment hints support users’ sensemaking by revealing certain conversational behaviors.
Low-level alignment hints can highlight forms of linguistic mimicry (e.g., strategic repetition of a partner’s wording) \cite{danescu2011mark}, while high-level alignment hints can reveal topic steering (e.g., reframing discussion topics to advance an agenda) \cite{clark1996using}.
Exposing these interactional patterns helps users interpret conversational dynamics, leading to modest gains in detection accuracy and confidence formation.
However, multi-level alignment-based hints provide richer interpretation by enabling users to reason across alignment levels.
In particular, scam-like strategies may involve patterns where low-level alignment scores remain high or continue to increase while high-level alignment scores decrease or remain low, signaling surface-level rapport alongside deeper divergence in intent.
\section{Conclusion and Future Work}
\label{sec:conclusion}

In this paper, we proposed multi-level alignment-based hints to support sensemaking in scam detection, capturing four different alignments across two levels, including low-level lexical and syntactic alignment and high-level semantic and situation-model alignment.
Our work treats scam risk as an evolving conversational dynamic, and embeds hints in a non-interruptive manner, thereby addressing two key limitations in previous work.
A preliminary evaluation reveals a consistent cross-level alignment pattern in real-life scam dialogues, where high-level alignment scores decline while low-level alignment scores remain stable as scam attempts approach.
User study results demonstrate that our proposal leads to more accurate scam detection (precision: 0.81, recall: 0.83, F1 score: 0.82) and earlier, more stable confidence formation over time compared with both keyword-triggered alerts and a no-hint baseline.
Ablation analysis further shows that the multi-level condition provides the strongest sensemaking support, outperforming single-level alignment variants.

Our future work has four potential directions.
First, because the current evaluation is limited to the LoveFraud02 dataset and thirty participants, future studies should examine whether similar alignment dynamics generalize across more diverse scam types and participant groups with a near-uniform age distribution.
Second, future user studies will involve qualitative analyses, such as interviews, to better understand user reasoning and examine HCI issues including potential over-reliance and possible misinterpretations.
Third, following a larger-scale user study, we will explore broader interaction paradigms and design principles for multi-level alignment-based hints.
Fourth, future work could explore the use of LLMs to generate structured, interpretable explanations of multi-level alignment dynamics, inspired by recent studies on transforming natural language artifacts into structured outputs \cite{li2024llm}.

\bibliographystyle{IEEEtran}
\bibliography{main}

\end{document}